\documentclass[preprint2]{aastex63}
\begin{document}
\title{The state of CO and CO$_2$ ices in the Kuiper belt as
seen by JWST}

\correspondingauthor{Michael Brown}
\email{mbrown@caltech.edu}
\author[0000-0002-8255-0545]{Michael E. Brown}
\affiliation{Division of Geological and Planetary Sciences\\
California Institute of Technology\\
Pasadena, CA 9125, USA}
\author[0000-0001-6680-6558]{Wesley C. Fraser} \affiliation{Herzberg Astronomy and Astrophysics Research Centre, National Research Council, 5071 W. Saanich Rd. Victoria, BC, V9E 2E7, Canada} \affiliation{Department of Physics and Astronomy, University of Victoria, Elliott Building, 3800 Finnerty Road, Victoria, BC V8P 5C2, Canada}

\begin{abstract}
JWST has shown that CO$_2$ and CO are common on
the surfaces of objects in the Kuiper belt and 
have apparent surface coverages even higher than
that of water ice, though water ice is expected to
be significantly more abundant in the bulk composition. 
Using full Mie scattering theory,
we show that the high abundance and the unusual spectral behaviour around the
4.26 $\mu$m $\nu_1$ band of CO$_2$ can be explained by a
surface covered in a few $\mu$m thick layer of $\sim 1-2$ $\mu$m
CO$_2$ particles. CO is unstable at the temperatures
in the Kuiper belt, so the CO must be trapped in
some more stable species. While hydrate clathrates { or 
amorphous water ice} are
often invoked as a trapping mechanism for outer solar system ices, the expected spectral shift of the absorption line 
for a CO hydrate clathrates or { trapping in amorphous ice} is not
seen, nor does the H$_2$O abundance appear to be high enough to explain the
depth of the CO absorption line. Instead, we suggest that the
CO is created via irradiation of CO$_2$ and trapped in the
CO$_2$ grains during
this process. The presence of a thin surface layer of CO$_2$
with embedded CO suggests volatile differentiation 
driving CO$_2$ from
the interior as a major process driving the surface
appearance of these mid-sized Kuiper belt objects,
{ but the mechanisms that control the small grain size and depth
of the surface layer remain unclear.}

\end{abstract}

\section{Introduction}
Observations from JWST have show that CO$_2$ and
CO ices are ubiquitous on the surfaces of
small bodies in the outer solar system
(Pinilla-Alonso et al., submitted,  H\'enault et al. 2023)
CO and CO$_2$ are the most abundant volatiles in comets after H$_2$O, at a $\sim$10\% mixing ratio, 
and presumably major
components of icy bodies in the outer solar system  \citep{2017RSPTA.37560252B}, but the difficulty of observing 
these species in their solid form means that little 
has been known of their 
distribution and abundance in the small body population today.

At the surface temperatures of the Kuiper belt and even throughout some of the Centaur range,
CO$_2$ should be stable, thus its presence in the
outer solar system should come as little surprise.
With its lower abundance than H$_2$O in comets, 
it should be expected to be a minor surface component 
on bodies wherever H$_2$O is exposed at the surface.
CO, in contrast, is unstable to sublimation 
anywhere inside $\sim$200 AU, and thus would not
be expected to appear on any small bodies in the
Kuiper belt \citep{1992acm..proc..545S}.

Here we use JWST spectra in the 0.8-5.2 $\mu$m 
range to examine the state and abundance of CO$_2$ and
of CO on objects in the Kuiper belt. We show
that, contrary to expectation, CO$_2$ appears
significantly more abundant on the surface of KBOs
than does water ice and that CO is retained at
temperatures significantly higher than its sublimation
temperature. We find that both of these properties can be explained
as a consequence of a thin layer of fine-grained CO$_2$ frost on these
objects.

\section{Observations and data reduction}
A large selection of KBOs and Centaurs is being observed by JWST program 2418. An overview of the observations
and first results is presented in Pinilla-Alonso et al. (submitted). 
CO$_2$ and CO appear in many of the spectra, and the 
spectral appearance of the CO$_2$ can be roughly 
sorted by the shape of the spectrum near the
4.26 $\mu$m $\nu_3$ absorption feature (H\'enault et al., 2023).
We 
examine two spectra, representative of the 
range of appearance of the CO$_2$ feature,
in detail to try to understand the
state and abundance of CO$_2$ and CO on these objects.
Table 1 lists the two objects and their known
characteristics, including diameter and albedo
from \citet{2012A&A...541A..94V} and \citet{2013A&A...555A..15F} and orbital elements
from JPL Horizons\footnote{ssd.jpl.nasa.gov}.

\begin{deluxetable*}{lcc}[t]
\caption{Properties of the observed KBOs}
\tablehead{\colhead{Name} & \colhead{2005 RN43} & \colhead{2002 TC302} }
\startdata
diameter (km) & 679$^{+55}_{-73}$ & 584$^{+106}_{-88}$\\
absolute magnitude (H)  & 3.89$\pm$0.05 &4.17$\pm$0.1  \\
albedo& 0.11$^{+.03}_{-.02}$ & 0.11$^{+0.05}_{-0.03}$ \\
semimajor axis (AU) & 41.8 & 55.7\\
eccentricity & 0.03 & 0.30\\
inclination (deg) & 19.2 & 35.0 \\
date of observation & 3 Nov 2022 & 2 Jan 2023 \\
heliocentric distance (AU) & 40.6 & 42.9\\
\enddata
\end{deluxetable*}

Each object is observed using the NIRSPEC integral field unit (IFU) mode with the prism
disperser \citep{2022A&A...661A..80J,2022A&A...661A..82B}, giving a spectral resolution of $R\sim 30-300$ from 0.6 to 5.2 $\mu$m and a spatial scale
of 0.1 arcseconds square within a 3 arcsecond square box. { Each of the KBOs appears as
a point source. }
We have experimented with multiple methods for
spectral extraction, and we find that a modification
of the standard JWST pipeline (version \#1.8.5) currently gives
the best results.
We begin our data analysis with the 
Level 2 ``rate'' files, which are the calibrated 
two-dimensional detector images showing individual
spectra from all 30 of the IFU slits. Examination of these files 
readily shows the effects of $1/f$ signal modulation caused by electronic
drifts during the detector readout. Many of the detector pixels are
not exposed to sky and so can be used to measure and remove this modulation.
We use the Level 2 ``cal'' files to find unilluminated pixels,
and, for each pixel in the ``rate'' image, we take the median of
all unilluminated pixels in a column within $\pm$150 rows to
define the modulation and subtract it from the ``rate'' image. 
The standard pipeline is then run on the new ``rate'' file, including
the AssignWcs,  MSAFlagOpen, SourceType, FlatField, PathLoss, and Photom
  steps. The final step in the standard pipeline is
the CubeBuild, which converts the two-dimensional spectral
images into three-dimensional spectral cubes.
We circumvent as much of the interpolation required for
mapping the slits to on-sky positions 
by forcing the pipeline to build the cubes with one axis
parallel to the slits and one perpendicular to the slits (using 
the ``internal\_cal'' geometry in CubeBuild). A single 
column of these cubes corresponds to a single slit. 

Significant background appears in the non-target parts of the spectral cubes,
particularly at the shortest and longest wavelengths. We remove this background
by taking the median of all of the pixels at each wavelength that are 5 pixels or
more distant from the center of the target. Dithered images could be used
for background subtraction, but the dither positions are often too closely spaced
to use effectively. 
Significant numbers of bad pixels appear in the image cubes, suggesting that PSF
fitting with robust outlier rejection could be a profitable method of spectral
extraction. Unfortunately, { fitting using
calculated PSFs (from, i.e., {\it WebbPSF}{\footnote{https://www.stsci.edu/jwst/science-planning/proposal-planning-toolbox/psf-simulation-tool}})
suffers from the undersampling of the PSF
by the detector,
even with these minimally resampled cubes. In these
data, the undersampling of the PSF}
causes spectral ripples in individual pixels of the image cube. The ripples are 
caused as the peak of the PSF slowly moves from the center of pixel, to the edge, to
an adjacent pixel as a function of wavelength. The frequency of these ripples 
thus corresponds to the frequency of the spectral trace moving from pixel-to-pixel,
which occurs once every $\sim$200 spectral pixels. The PSF as sampled by the
detector thus also changes over this pixel scale in a manner not 
currently captured by any of the available PSF modeling tools. We circumvent this
difficulty by constructing an empirical PSF at each wavelength from the median
of the spectral images within 25 pixels of that wavelength normalized to a total value
if unity. We then divide the
spectral image by the empirical PSF at that wavelength and take the median 
of this division { within a 3 pixel radius} as the brightness at that pixel. { The final spectrum 
is thus free of the pixel-to-pixel ripples introduced
by the PSF undersampling.}

To obtain the relative reflectance of the object, we divide the extracted spectrum
by an identically reduced spectrum of a PRISM observation of the solar analog
2MASS J16194609+5534178 \citep{2022AJ....163..267G} from  Program \# 1128.
This division removes any wavelength dependent calibration issues in the JWST pipeline,
corrects for the wavelength dependent PSF changes,
and converts the flux to a relative reflectance. 
{ We normalize each spectrum to have a median value of 
unity between 1.5 and 2.5 $\mu$m.}
The final spectra of our two example objects 
are shown in Figure 1.

\begin{figure}
\includegraphics[width=3.7in]{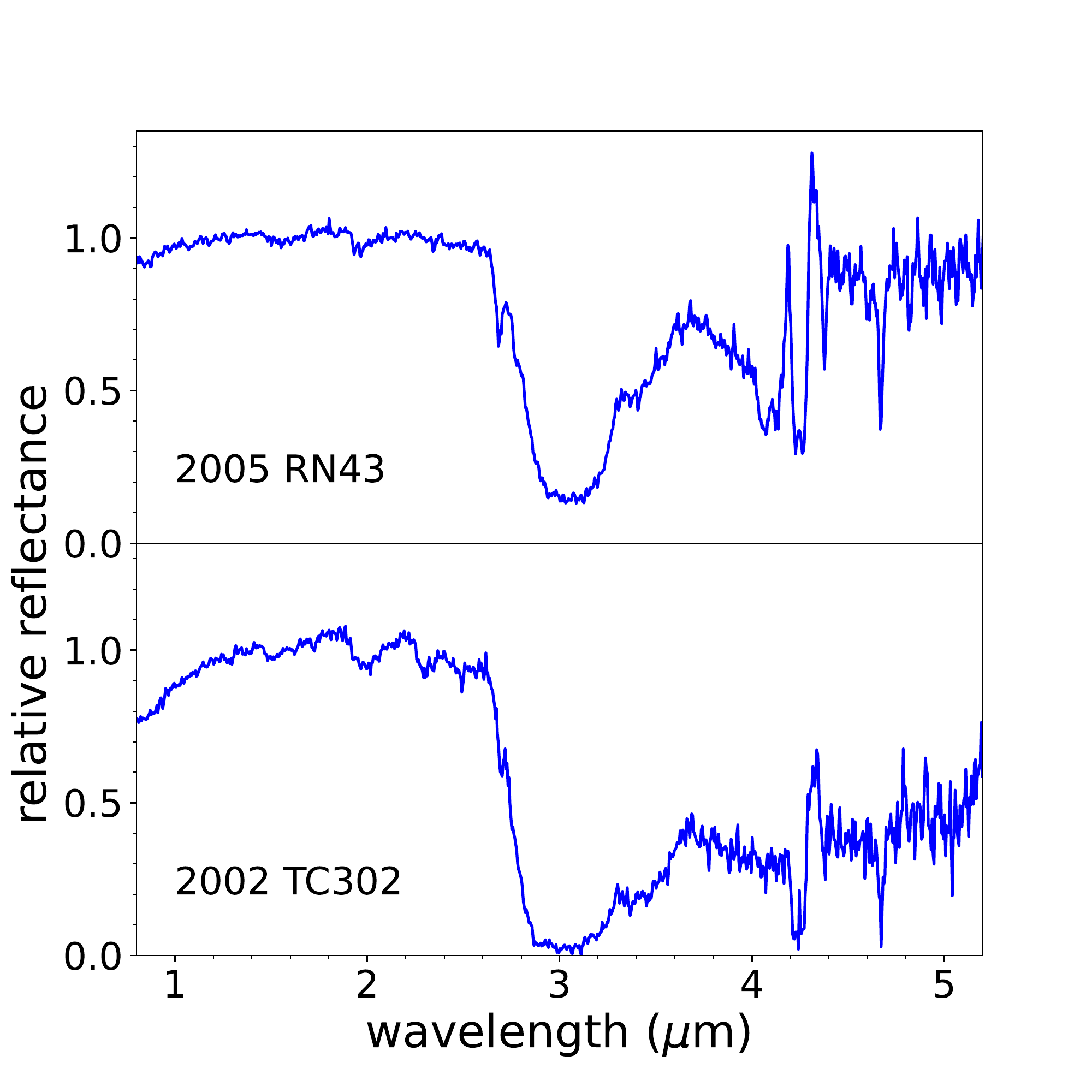}
\caption{The spectra of two representative KBOs
from the sample. Each shows subtle 1.5 and 2.0 $\mu$ 
absorption features indicative of a modest exposure of
water ice, deep 3 $\mu$m absorption consistent with 
a tholin-like processed organic surface, a broad
3.4-3.6$\mu$m absorption owing to organic materials,
a narrow 4.6 $\mu$m absorption from CO, and numerous
and variable features from 4.0 to 4.4 $\mu$m all attributably to CO$_2$. }
\end{figure}

\section{Results and modeling}
The features surrounding the deep 4.26$\mu$m CO$_2$
absorption on 2005 RN43 { and 2002 TC302}
are unlike any reflectance 
spectrum previously seen. A 4.26$\mu$m absorption
has been seen on the icy satellites of Jupiter
\citep{2003JGRE..108.5036H} and Saturn \citep{2010Icar..206..561C}, but
in all cases the absorption is modest
and shows no additional absorptions nearby. In these
cases, however, CO$_2$ is not in pure form but must
be trapped as a minor constituent in a background 
material to allow the CO$_2$ to be stable at the 
elevated temperatures of these satellites. These { new KBO} 
observations are the first at temperatures where
pure CO$_2$ is stable and potentially dominating
the spectral region around the 4.26$\mu$m line.
We thus carefully examine the reflectivity of
pure CO$_2$ through this region.

Between 4.0 and 4.5 $\mu$m, CO$_2$ has not just
a large change in absorptivity, but its real 
index of refraction goes from its typical value
of $\sim$1.3 at shorter wavelengths,
approaches zero near 4.19$\mu$m, climbs above a 
value of 5 at 4.27$\mu$m and then drops to 1.5
at longer wavelengths \citep{2020ApJ...901...52G}. Dramatic
changes such as these in the real index of refraction
alongside large changes in the imaginary index of
refraction (related to absorption) cause complicated
spectral behaviour. In Figure 2 we plot the spectrum of
both objects, along with the real and imaginary indices
of refraction. As can be seen, the most anomalous-appearing
spectral behaviour of 2005 RN43 appears at
precisely the locations where the real index of
refraction goes through unity, where it approaches zero,
where the imaginary index peaks, and where the imaginary
index falls back to zero but the real index changes
rapidly. 
\begin{figure}
    \centering
    \includegraphics[width=3.7in]{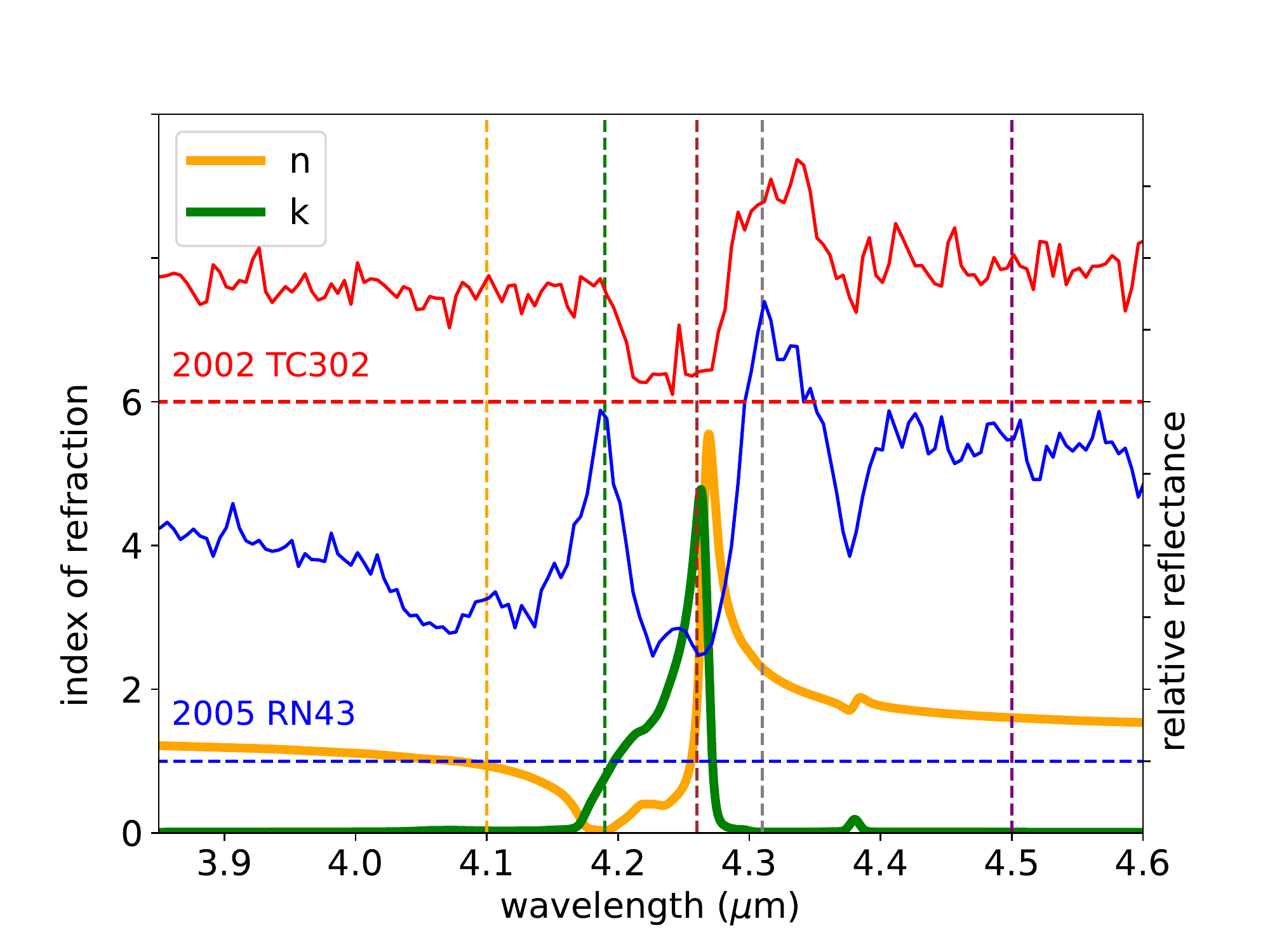}
    \caption{The real ($n$) and imaginary ($k$) indecies of 
    refraction of CO$_2$ compared to the spectra
    of 2005 RN43 and 2002 TC302, which are both offset for clarity with their zero points shown as blue and red horizontal dashed lines, respectively. Major spectral features on both objects
    occur at the locations marked by dashed lines,
    including when the real index passes through unity,
    when it approaches zero, where the imaginary
    index peaks, and where the imaginary index
    returns to near zero while the real index
    changes rapidly.}
    \label{fig:index}
\end{figure}

Because of the complexity of the behaviour in this region,
we resort to full Mie theory to understand the
reflectivty across this wavelength region. We
use the Python package {\it miepython}, which implements
and has been validated against the \citet{1979msca.rept.....W} algorithm, to calculate the single-scattering
albedo of spherical CO$_2$ particles from 
very small 0.1 $\mu$m particles, in the Rayleigh scattering 
regime, to 100 $\mu$m particles, in the geometric
optics regime. Using the single-scattering albedo,
we calculate the geometric albedo assuming a
uniform surface of isotropic scatters using
the formulation of \citet{1993tres.book.....H}, equation (10.37). 

{ Mie scattering calculations are only strictly accurate 
for well-separated spheres, which is clearly not the case on
a planetary surface. \citet{1997Icar..125..145M} investigated
the spectra of olivine and quartz 
powders with grain sizes near and below the wavelength 
of spectral features and found that a combined
Mie theory/Hapke modeling approach as used above provides the best modeling
match to laboratory data. Some limitations to this approach are discussed below.
}

Figure 3 shows the geometric albedo
as a function of the particle size of the scatterers
for the 5 key wavelengths shown in Figure 2. One
point is immediately apparent from this Figure. For the geometric
albedo at 4.31 $\mu$m to rise above that at the 4.19 $\mu$m peak or
at the 4.5$\mu$m continuum beyond that, the grains must be smaller
than about 2 $\mu$m. It is also clear that as particles
approach the Rayleigh scattering regime the contrast between the
peak at 4.19 $\mu$m and the trough at 4.26$\mu$m becomes increasingly
smaller while the overall albedo declines quickly. 
\begin{figure}
    \centering
    \includegraphics[width=3.7in]{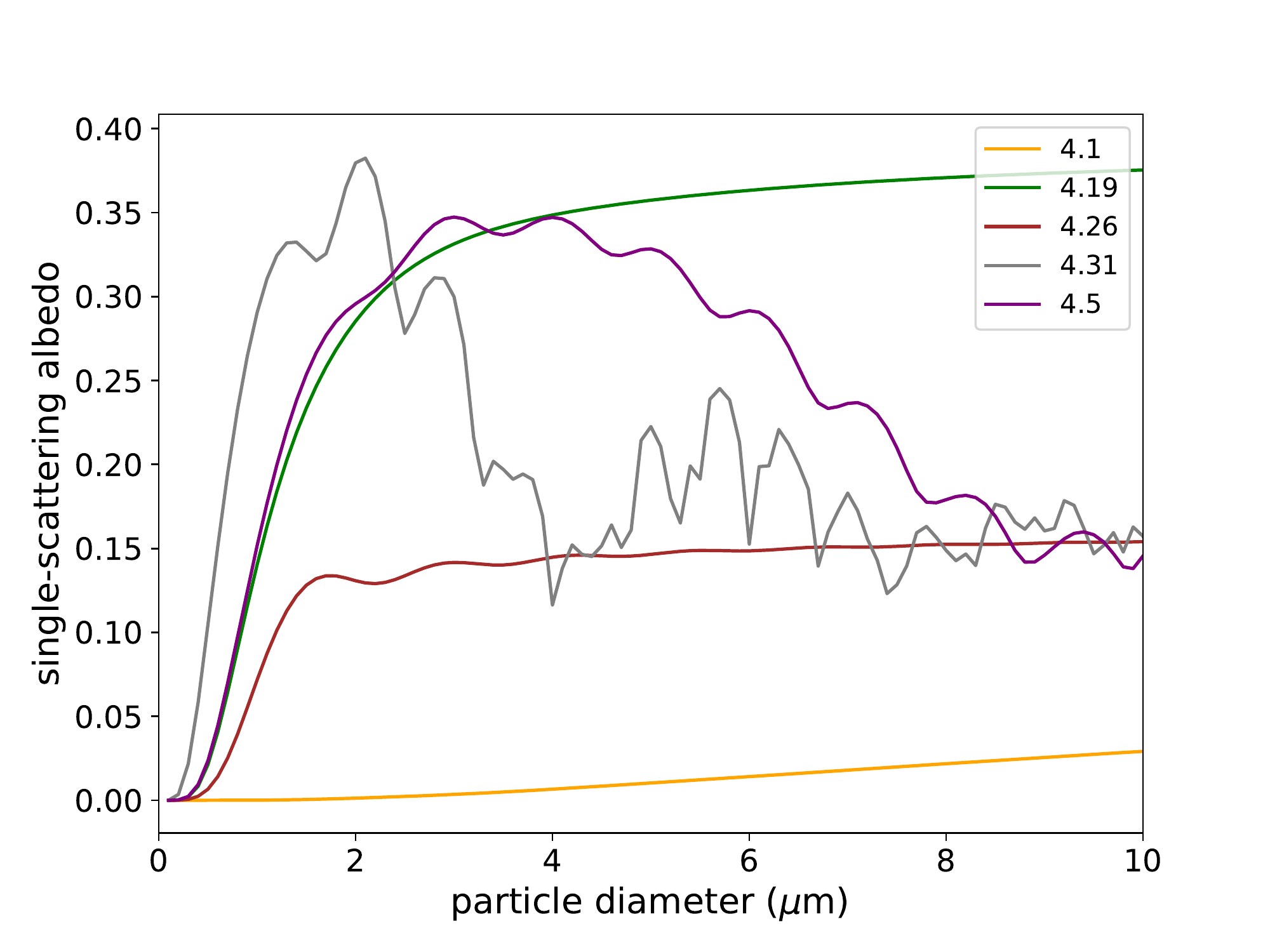}
    \caption{The geometric albedo from a collection
    of Mie scattering particles from 0.1 to 10 $\mu$m in
    diameter. The colors of the lines correspond to
    the scattering behavior 
    at the different wavelengths
    of the same
    colors of the dashed lines in Figure 2. The strong changes in both the real and imaginary index of refraction across this spectral region makes the spectral behaviour an extremely complicated and sensitive function of particle size.}
    \label{fig:mie}
\end{figure}
To understand the full spectrum through this region we show, in Figure 4,
model spectra of a uniform isotropic surface of Mie scattering particles of 
various sizes. For 0.5 $\mu$m particles the 4.26$\mu$m absorption no
longer appears distinct, while at sizes of 1-2 $\mu$m the doublet in that
absorption band -- which is clear in the data -- emerges. We conclude that the
main structures seen in the spectrum of 2005 RN43 around 4.26 $\mu$m, namely
the absorption near 4.10 $\mu$m, the narrow peak at 4.19 $\mu$m, the doublet
absorption around 4.26 $\mu$m, and the broader strong peak at 4.31 $\mu$m can
all be explained if this object is covered in CO$_2$ grains between about 1 and 2 $\mu$m 
in diameter. { The behaviour seen in 2002 TC302, with a single
peak redward of the deep CO$_2$ absorption is not seen in
these models.}
\begin{figure}
    \centering
    \includegraphics[width=3.7in]{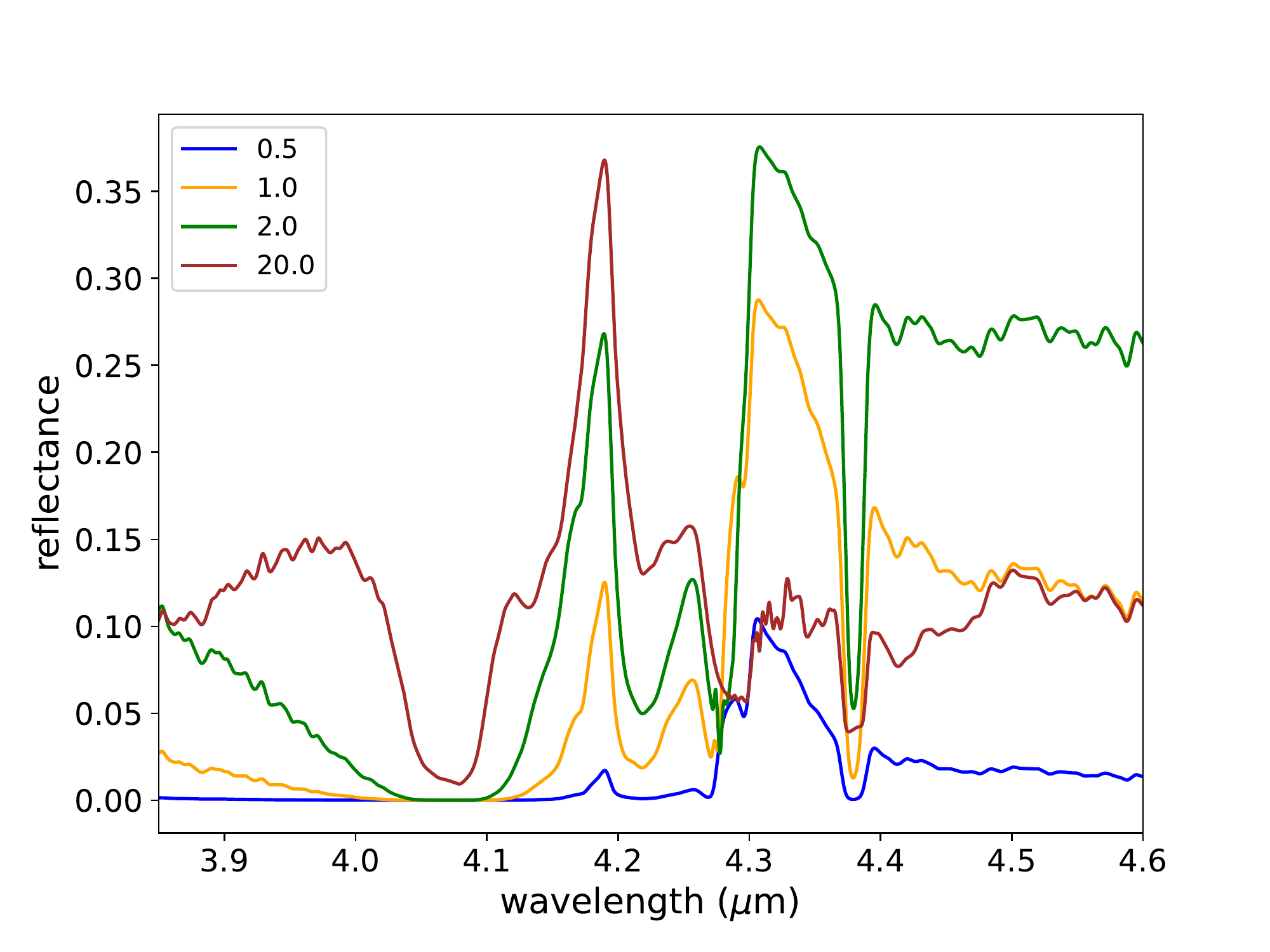}
    \caption{Full spectral models of a collection
    of Mie-scattering CO$_2$ particles of 4 particle
    sizes. The prominent peaks at both 4.19 and 4.34 $\mu$m are only seen for particle diameters of
    1-2 $\mu$m}
    \label{fig:specmod}
\end{figure}

One feature prevalent in all of the model spectra with small CO$_2$ grains is that
the albedo is a minimum near 4.1 $\mu$m, where the real index of refraction passes through 
unity. The imaginary index of refraction is small but non-zero at this wavelength, so 
light will pass through all grains unaffected by refraction or reflection, and simply be
absorbed after passage through a large number of grains. The albedo will thus stay near
zero. In the real spectra of the KBOs observed in the full program, however, the albedo is always lowest in the 
4.23/4.26 $\mu$m absorption doublet (H\'enault et al., 2023). Such a spectral shape can be a simple consequence
of reflection from a layer of CO$_2$ particles covering a reflective medium. At 4.1 $\mu$m the
CO$_2$ is essentially transparent, so it will propagate through the CO$_2$ layer and reflect
off of the underlying surface. At 4.23/4.26 $\mu$m the light can -- depending on the optical
depth of the layer -- still be fully absorbed within the CO$_2$ layer, leading to deeper
absorption lines. We model this behaviour using the formalism for layered media of \citet{1993tres.book.....H} (equation 9.14). We assume a thin layer of CO$_2$ particles with a depth of $l$ and particle
diameter of $d$ overlying a surface with a constant geometric albedo of $a$. Figure 5 shows
three spectral models with only small changes in these parameters. These models reproduce
the basic behaviour of both { the 2005 RN43-type of CO$_2$ spectrum, with sharp lines on either side of the main absorption,
and the 2002 TC302-type spectrum, with a single line redward of
the CO$_2$ absorption.} In the models shown here, the
CO$_2$ layer is between 1 (orange and green lines)
and 6 (blue line) particle diameters thick.

{ The contrast between the continuum and
the peaks seen in these modeled spectra is higher than 
that seen in the real KBOs. Interestingly, the same
behaviour is seen by \citet{1997Icar..125..145M} in the
laboratory vs. modeled spectra of quartz, which has 
similar values of the both the real and imaginary index
of refraction in the 9 $\mu$m region as CO$_2$ does in the 4.3 $\mu$m region. We thus suspect that the too-high contrast of the
models is a consequence of the difficulties of using Mie 
theory to model tight packed grains such as these and 
that we are indeed observing a surface covered with such small
grains.}

\begin{figure}
    \centering
    \includegraphics[width=3.7in]{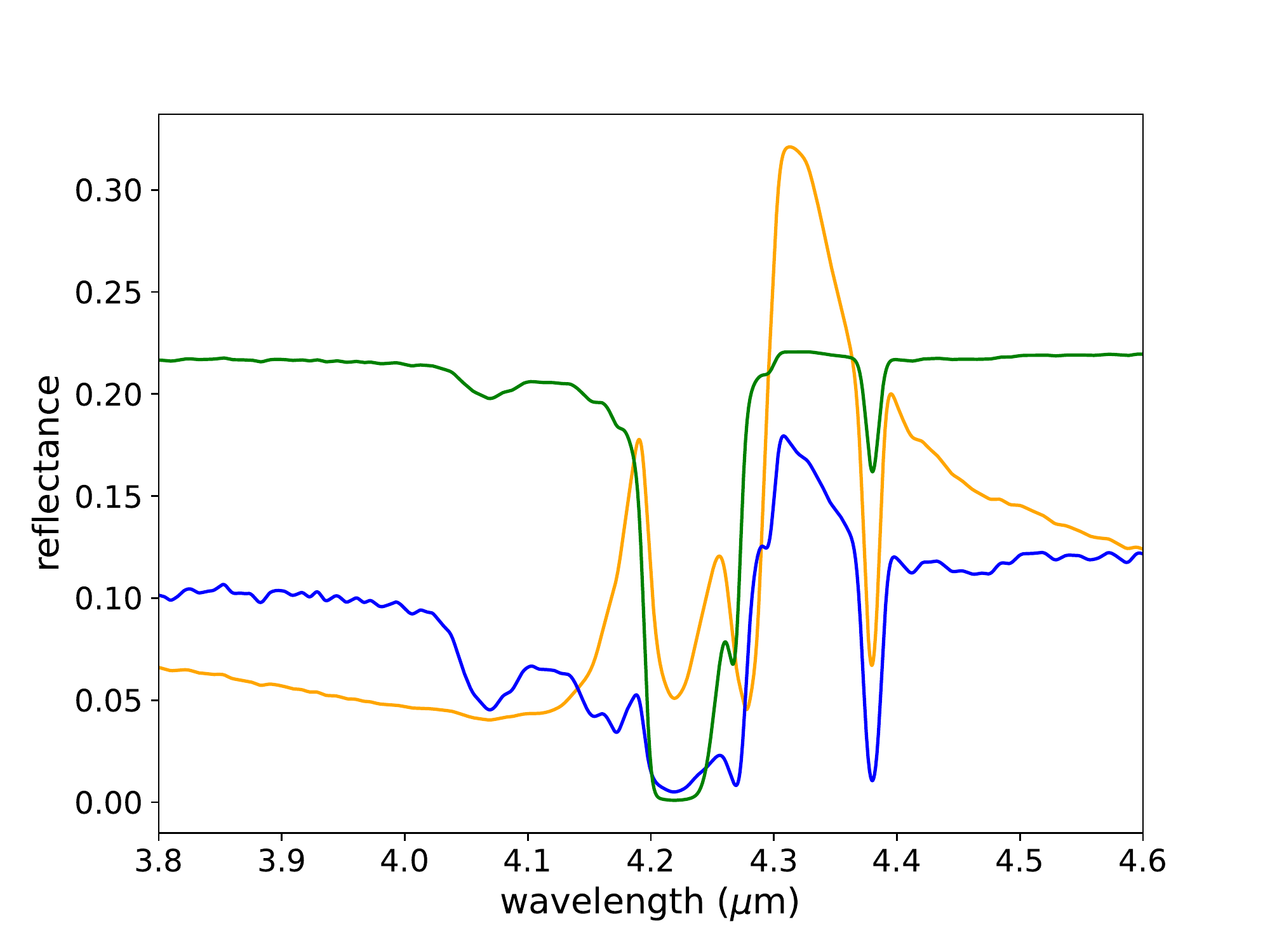}
    \caption{Spectral models of a thin layer of CO$_2$
    particles overlying a reflective medium. Models
    such as these are needed to correctly reproduce
    the non-zero albedo at 4.1 $\mu$m where CO$_2$ becomes essentially transparent. Slight changes
    in grain sizes and CO$_2$ layer thickness give
    rise to many different apparent spectral types,
    allowing the full range of CO$_2$ detected in
    the Kuiper belt to be explained.The three
    models shown have background albedo,
    CO$_2$ grain diameter, and CO$_2$ layer thickness of 0.05, 2$\mu$m, 2$\mu$m (orange); 0.16, 0.7$\mu$m, 10 $\mu$m (blue); and 0.23, 0.5$\mu$m, 1$\mu$m (green).}
    \label{fig:layer}
\end{figure}

An interesting consequence of the highly variable nature of the spectra from 4.0
to 4.4 $\mu$m is the difficulty of quantitatively interpreting the $^{13}$CO$_2$ line
at 4.38 $\mu$m. In Figure 5, for example, the $^{13}$CO$_2$ abundance is just
the terrestrial value and is seen in the measured values of $n$ and $k$ simply
because of natural contamination of the laboratory sample of CO$_2$. In
each modeled spectrum, however, though the $^{12}$C/$^{13}$C
ratio
is identical, the depth of the $^{13}$CO$_2$ line is
variable. The variability of this line in these models
highlights the difficulty of the using these observations
to make precise quantitative statements
about the $^{12}$C/$^{13}$C ratio in these objects. 
Nonetheless, the observations of the $^{13}$CO$_2$
line on these objects does not appear strongly
inconsistent with the strength seen in the spectral
models, suggesting that the isotopic ratio 
is not strongly inconsistent with the terrestrial
value of 89. Such a finding is not surprising given the
measurements of the $^{12}$C/$^{13}$C in cometary
C$_2$, CN, and HCN, which also give values close to the terrestrial value
\citep{2015SSRv..197...47B}.

\section{CO}
The presence of CO on small bodies in the Kuiper belt
requires that the CO be trapped to prevent sublimation
at these temperatures. Trapping of CO { in amorphous water ice or as
a clathrate hydrate}
is often invoked as a possible process
important in the outer solar system
\citep{1985Icar...63..317B,2005SSRv..116...25G}. The $\nu_1$
absorption feature of enclathrated
CO is spectrally shifted from its position at 4.675 $\mu$m in
pure CO ice \citep{2011Icar..212..950D}. 
In Figure 6 we show a detail of the
spectrum of 2005 RN43 near the CO region. 

Measuring the precise wavelength of the CO feature
is complicated by the uncertainty that still remains in
the wavelength scale of the PRISM mode observations. 
To understand possible wavelength uncertainties, we examined
wavelengths of known emission lines in a star forming region
of NGC 7319 (Program 2732). After correcting
for the redshift, we find that beyond 4 $\mu$m, the wavelength
scale is off by about half a pixel ($\sim$0.003 $\mu$m)
and that the
wavelength scale between 4 and 5 $\mu$m likewise changes
by about half a pixel.
To account for his effect as best as possible, 
we set the wavelength scale by forcing the $^{13}$CO$_2$
absorption to occur precisely at 4.381 $\mu$m (which involves a 0.005 $\mu$m 
shift from the nominal wavelength solution). While trapped CO$_2$ will have a wavelength shift from this nominal wavelength \citep[for example on Ganymede;][]{2003JGRE..108.5036H}, the high abundance of CO$_2$ 
ensures that we are not seeing a small trapped population here. Thus, based on the
analysis of NGC 7319, the wavelength
solution at the location of the CO line should be accurate 
to within at least a 0.005 $\mu$m pixel.

With this wavelength solution, the CO line on 2005 RN43
appears 
at the 4.675 $\mu$m wavelength of pure CO, and not at the 
4.686 $\mu$m wavelength of
CO clathrate nor the 4.682 $\mu$m wavelength measured for trapping on amorphous water ice at 50 K \citep{1988Icar...76..201S}. This 0.007 - 0.011 $\mu$m difference is well above any
expected wavelength uncertainty.
Lack of evidence for a water ice CO clathrate or trapping in amorphous water ice is perhaps not surprising given
the weak 1.5 and 2.0 $\mu$m absorption lines of water ice. 

Given the large surface coverage of CO$_2$ on these objects, we consider CO
trapped in this species. \citet{2019ApJ...883...21S} examined the trapping of CO in CO$_2$ and 
found that such trapping can be even more efficient than trapping in H$_2$O. The trapped CO was stable at temperatures as high as 80 K, even higher
than the expected surface temperatures of these objects. At 50K, the wavelength of
the absorption from the trapped CO is 4.673 $
\mu$m, consistent with the position of the line
seen here. While the wavelength match is
good, the mechanism of gas trapping used in these
experiments -- mechanical mixing of the gasses before
freezing onto a substrate at 17 K -- does not 
obviously apply to these objects.

A more promising CO trapping mechanism involves both the production and trapping of
CO in CO$_2$ through irradiation. \citet{2015A&A...584A..14M} showed the UV irradiation of CO$_2$ creates
CO which remains stable even at temperatures as high as 90 K. The trapping mechanism here
is different from either the enclathration or the physical trapping discussed above,
so it is possible that the wavelength shift of CO trapped in this manner is different 
from those mechanisms. Unfortunately, the wavelength of the CO absorption at 
temperatures relevant for the Kuiper belt was not measured. Physically and chemically,
this explanation for the presence of CO at elevated temperatures seems most promising, but laboratory measurement of the 
wavelength of CO created and trapped at KBO-like temperatures
is clearly needed.
\begin{figure}
    \centering
    \includegraphics[width=3.7in]{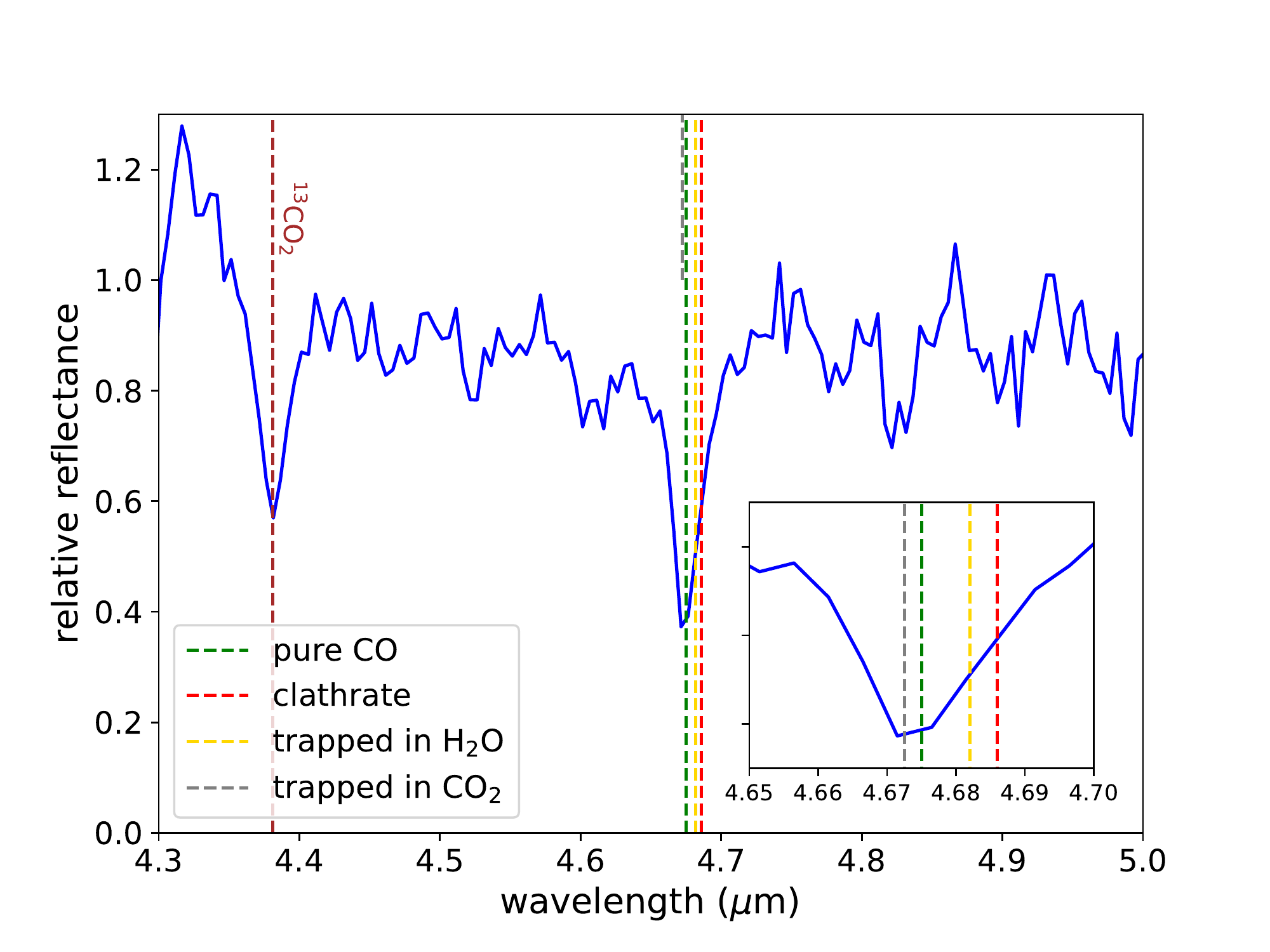}
    \caption{A small portion of the spectrum of 2005 RN43 showing both the $^{13}$CO$_2$ line, which
    we use as a wavelength reference, and the CO line.
    The wavelength of the CO does not match that
    of CO clathrate hydrates or CO trapped in amorphous ice, but is a close match
    to pure CO (which is unstable at this temperature),
    and also to CO trapped in CO$_2$ ice.}
    \label{fig:CO}
\end{figure}

\section{Discussion}
The presence of a thin layer of fine-grained CO$_2$ 
appears common on objects in the Kuiper belt. CO is likewise
common and must be trapped within some thermally stable material. These two observations lead to a natural model
for the presence of both of these molecules. 
We posit that CO$_2$ from the interior of these objects is heated and diffuses 
toward the surface where it is retained as a fine-grained frost. Irradiation
of the CO$_2$ (UV, solar wind, and/or cosmic ray) then produces CO within these CO$_2$ grains. 

{ This simple idea hides some difficult
complexities.
Many of the other KBOs in this program are 
seen to have CO$_2$ features
manifesting like the two discussed 
here (H\'enault et al., 2023).
Given 
the high level of sensitivity of the spectral models
to the grain size and layer thickness, the lack
of a wider variety of spectral types is surprising. \citet{1993tres.book.....H} argues that for
grain sizes much smaller than the wavelength of the
light the 
scattered radiance tends to sample inhomogeneities in
the medium that are of order of the wavelength divided
by pi, which is 1.35 $\mu$m in our case and remarkably close
to the preferred range of particle size solutions that we find.
In this case the observations would be indicative of grains
a few microns or smaller without the need to be a specific size,
at least partially alleviating the fine-tuning problem.

The thickness of the CO$_2$ layer also exerts a strong 
control on the modeled spectrum, in particular in the albedo
at 4.1 $\mu$m where $n=1$ and CO$_2$ is transparent.
Layers larger than a few microns begin to look more like the
pure CO$_2$ spectra of Figure 4 and are hard to reconcile
with the observations. The need to use thin layers
in the models is no obvious consequence of decifiencies
of Mie or Hapke
theory, so we believe that such thin layers are indeed
intrinsic to the KBOs. How such thin layers
are formed and preserved with such a small thickness range
is unclear, but the tight constraints suggest 
an ongoing balance between formation and destruction that must
be common to these objects.}

This process of { outward volatile transport}
should operate most strongly on the
larger KBOs which have received the most heating. The size range
probed in this program is modest, so smaller objects will
have to be observed to see if this prediction is borne out.
No modeling of interior volatile evolution on objects of these sizes has yet been attempted, though
for smaller sizes CO$_2$ is thought to remain unchanged by the smaller amount of interior heating \citep{2001AJ....121.2792D}. 
Understanding the implications of this new body of
data will be a critical next step in piecing together the
chemical history of the outer solar system.

\acknowledgements
We would like to thank John Stansberry and 2 anonymous reviewers
for interesting comments on the manuscript.
This work is based on observations made with the NASA/ESA/CSA James Webb Space Telescope. The data were obtained from the Mikulski Archive for Space Telescopes at the Space Telescope Science Institute, which is operated by the Association of Universities for Research in Astronomy, Inc., under NASA contract NAS 5-03127 for JWST. These observations are associated with program \# 2418
and the specific observations analyzed can be accessed via \dataset[DOI: 10.17909/tnkm-hq35]{https://doi.org/10.17909/tnkm-hq35}.
The authors acknowledge the team led by PI N. Pinilla-Alonso for developing their observing program with a zero-exclusive-access period.

\facility{JWST (NIRSPEC)}
\software{miepython \\({http://miepython.readthedocs.io})} 
\

\

\

\

\

\

\

\

\

\eject
\bibliographystyle{aasjournal}

\end{document}